\documentclass[10pt,twocolumn,preprintnumbers,amsmath,amssymb,prb]{revtex4}
\usepackage[dvips,bookmarks=false]{hyperref}
\usepackage{epsfig} 	
\usepackage{epsf}
\usepackage{dcolumn}
\usepackage{bm}

\usepackage[latin1]{inputenc}
\usepackage{amsmath}
\usepackage{latexsym}
\usepackage{notebook2e}

\def\dsize{\textstyle}
\def\tst {\textstyle}
\def\dst #1{\displaystyle #1}
\def\MATH#1 {\begin{math}{\dst #1}\end{math} } 
\def\EQ#1 {\begin{equation}{\begin{array}{ll} \dst #1 \end{array}}\end{equation}}
\def\ft#1 {{\overvar{#1}{{}_{_{\dsize\hat{}}}}}}
\def\bs#1 {\boldsymbol #1}

\def\bold#1{\mbox{\bfseries #1}}
\def\mb#1{{\bold #1}}
\def\vb#1{\vec{\mb{#1}}}
\def\sca{{a}}
\def\sce{{e}}
\def\scs{{s}}
\def\sck{{k}}

\def\v#1{\vb{v}_{\rm #1}}
\def\u#1{\vb{u}_{\rm #1}}
\def\m#1{m_{\rm #1}}
\def\V{\vb{V}_{{\rm cm}}}
\def\q{\theta}
\def\cq{\vartheta}
\def\R{\vb{R}_{{\rm cm}}}
\def\e#1{\vb{e}_{#1}}
\def\r#1{\vb{r}_{#1}}
\def\dprime{{\,\prime\prime}}
\def\arctan{\tan^{-1}}
\def\arccos{\cos^{-1}}
\def\d{\dsd}
\def\Uo{U_{\rm o}}
\def\rmin{\scr_{p}}

\renewcommand \( {\left(}
\renewcommand  \)  {\right)}
\newcommand{\csep}[2][1.4]{\renewcommand{\arraystretch}{#1}\setlength\arraycolsep{#2 bp}}

\begin{document}

\title{Binary Collision Orbits and the Slingshot Effect}

\author{Amaro J. Rica da Silva}
\email{amaro@fisica.ist.utl.pt}
\author{José P. S. Lemos}
\email{lemos@fisica.ist.utl.pt}
\affiliation{Centro Multidisciplinar de Astrofísica-CENTRA \\\&\\ Physics Dept., Instituto Superior Técnico, Universidade Técnica Lisboa\\ Av. Rovisco Pais, 1049-001 Lisbon, PORTUGAL}

 \begin{abstract}
We derive the equations for the gravity assist manoeuvre in the general 2D case without the constraints of circular planetary orbits or widely different masses as assumed by Broucke\citep{Bro88}, and obtain the slingshot conditions and maximum energy gain for arbitrary mass ratios of two colliding rigid bodies.
Using the geometric view developed in an earlier paper by the authors\citep{RicadaSilva06} the possible trajectories are computed for both attractive or repulsive interactions yielding a further insight on the slingshot mechanics and its parametrization\citep{AJRS}. 
The general slingshot manoeuvre for arbitrary masses is explained as a particular case of the possible outcomes  of attractive or repulsive binary collisions, and the correlation between asymptotic information and orbital parameters is obtained in general. 
\end{abstract}

\maketitle

\section{Introduction}

The slingshot or gravity assist manoeuvre\,\citep{Bro88}${}^{-}$\nocite{Dykla}\nocite{KJE05}\nocite{Longuski91}\citep{RCJ03} is often considered as part of a restricted three-body problem and its use has been associated in the literature mostly with spaceflight strategies \citep{RACCA}${}^{-}$\ \nocite{Malyshev03}\citep{Ocampo03} with some applications in astrophysics for the study of mass ejection from binary clusters \citep{Saslaw74} and the proposal of new General Relativity tests \citep{Longuski01}. 
In reality, the design of spacecraft trajectories between two planets is a many-body problem except for the slingshot part, which is in most designs well approximated by an elastic binary collision.
This work focuses on the slingshot  manoeuvre as a particular case of a general binary elastic collision between massive objects subjected to central interaction forces. 
In a previous work \citep{RicadaSilva06} the geometric determination of binary collisions was introduced, and the possible outcomes were in some cases surprising \citep{AJRS}. 
We have obtained a parametrization of all possible outcomes of a binary elastic collision in an arbitrary frame, and from the mass ratios and initial velocities as asymptotic initial conditions we obtain a picture not only of the final asymptotic velocities in terms of a single parameter $\q$ in the 2D case, but also the detailed description of the two-body motion that fits these asymptotic data in the case of the gravitational or coulombian interaction. 
We are therefore in condition to determine which precise orbital parameters must be chosen to obtain a desired effect on a flyby of a satellite about a planet or star, be it a gravity assisted boost or capture.
The conditions for a gravity-assisted manoeuvre of a satellite are often loosely associated to a flyby in front or behind the planet.\citep{Bro88}  
Both in the case of attractive as well as repulsive collision forces,  a harder look must be performed to really grasp what the critical ingredient is. In particular we show how the geometry and the timing of arrival at the point of closest approach (periapsis), to wit the relative position of the bodies with respect to the normal to the Center of Mass velocity $\V$ at that point, determines the outcome of the collision and refine the phenomenological rule-of-thumb  that a flyby in front of the planet results in a slowing of the satellite whilst a flyby behind the planet's trajectory would result in a boost.\cite{LPS} 
In fact, the asymptotic description of a collision is somewhat elusive in this respect. By scaling out the interaction in all space and time dimensions there is naturally a loss of information of precisely where and when are the two bodies for given initial velocities, so a determination must be made as to what corresponds in a real problem to these times and positions. It could be claimed that these initial velocities should be those of the bodies when their distance is equal to the sum of the radii of their respective spheres of influence\citep{Barger}. 
But since this too is a fuzzy concept this is not much of an improvement. In fact, that information is only present when enough conditions are specified to determine the collision outcome uniquely.\cite{Asada07}
Thus, in the 2-dimensional case, the circumference of possibilities for the velocity outcomes of one of the bodies encodes the  missing information about where the bodies initially are simultaneously when they have the given velocities. 
This can also be translated into an impact parameter in the non-inertial body-frame for one of the masses (or reduced-mass frame) but that begs the question of viewing the collision in the laboratory frame. 
In this paper the assumption will therefore be made that at $t=0$ the Center of Mass (CM) will be at the origin of the laboratory frame (LF).

\begin{figure*}[ht]
\hspace{-2em}\includegraphics[scale=.70]{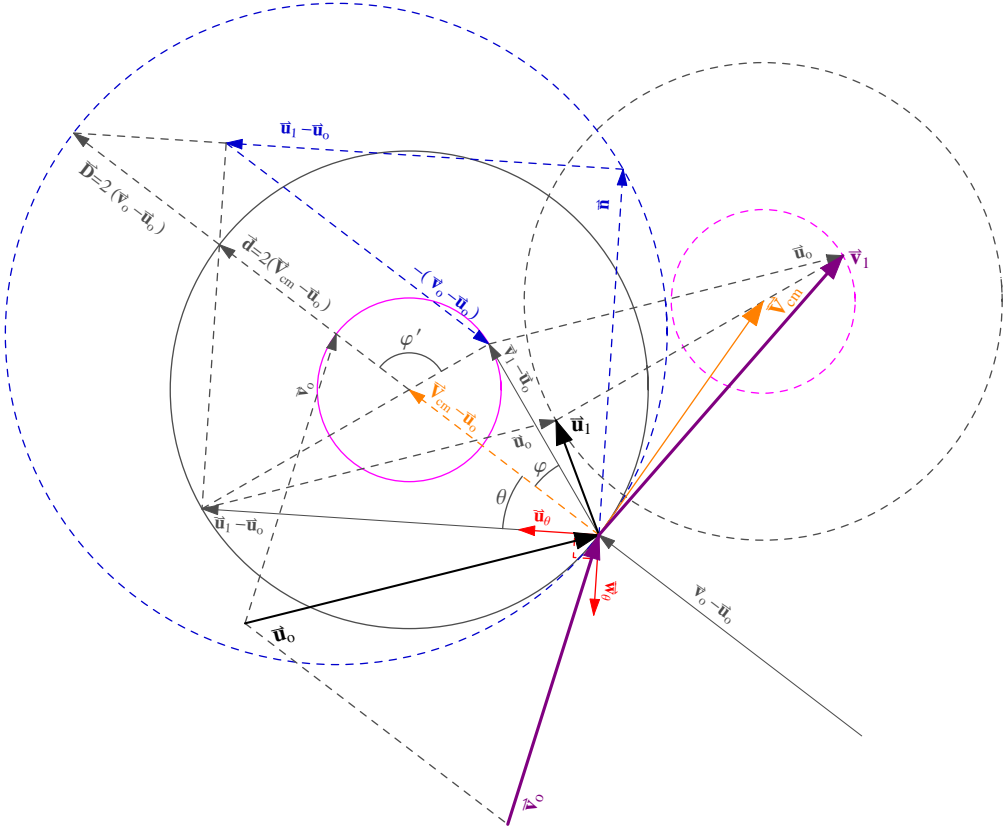}
\caption{{ \hsize=.8\hsize Collision diagram for different masses $\m{v}>\m{u}$  with initial velocities $\vec{\bold v}_{\rm o},\vec{\bold u}_{\rm o}$, and its relation with the collision diagram as seen from the $u$-body (mass $\m{u}$) initial rest frame $\scS_{u_{\rm o}}$. In this frame the circumference with diameter $\vec{\bold d}=2(\V-\vec{\bold u}_{\rm o})$ is the locus of all possible outcomes $\vec{\bold u}_1-\vec{\bold u}_{\rm o}$ parametrized by the scattering angle $\q$. A choice of $\q$ determines $\vec{\bold u}_1-\vec{\bold u}_{\rm o}$ and an orthogonal vector $\vb{n}$ restricted to a circumference of diameter $\vec{\bold D}=2(\vec{\bold v}_{\rm o}-\vec{\bold u}_{\rm o})$. The outbound velocity $\vec{\bold v}_1-\vec{\bold u}_{\rm o}$ is geometrically defined as $\vec{\bold n}+\vec{\bold u}_1-\vec{\bold v}_{\rm o}$ and lies on a circumference centered on $\V-\vec{\bold u}_{\rm o}$ with radius $|\vec{\bold v}_{\rm o}-\V |$. The angles $\q$ and $\varphi$ represent the $u$- and $v$-scattering angles relative to the incoming velocity of the CM in the $u$-body initial rest frame and $\varphi^{\prime}$ is the scattering angle in the CM frame.}}\label{fig:GEN}\vspace*{5pt}
\end{figure*}

Diagrams like the one in Fig. \ref{fig:GEN}, introduced in a previous work\citep{RicadaSilva06}, are used to correlate the asymptotic information, provided by initial velocities far away from the periapsis, with the possible eccentricities, focal distances and other orbital parameters for open Keplerian orbits in case of gravitational attractive or Coulombian repulsive scattering. 
These diagrams depict the relation between incoming laboratory frame asymptotic velocities $\v{\rm o}$ and $\u{\rm o}$, of masses $\m{v}$ and $\m{u}$ respectively, and their final asymptotic velocities $\v{1}$ and $\u{1}$, through a computation involving the   scattering angle $\theta$ of  the $\m{u}$ mass, measured  in its initial asymptotic $u$-body  frame from the direction of the incoming relative velocity $\v{o}-\u{o}$ of the $\m{v}$ mass. In that $u$-body frame the circumferences of possible velocity outcomes are easily drawn and their image in the laboratory frame can easily be deduced, thus yielding information about the possible directions and magnitudes of asymptotic final velocities $\v{1}$ and $\u{1}$ for both masses. 
In this way the orbits can be viewed in the laboratory frame and a study can be made, for instance, for the optimal incidence angle on a planetary fly-by that delivers the maximum allowed velocity boost in a chosen direction. 
The energy gain is obtained both in the case of idealized point particle collisions and extended object collisions, where the periapsis distance is constrained by a minimum value below which the collision is no longer elastic. 
The relation of this asymptotic information with the actual trajectory can also be displayed using the fact that the reduced-mass-frame trajectory axis and asymptotic directions are already included in these diagrams, and thus also all the orbital parameters can be deduced or introduced here.  
\medspace
The final velocities can be determined from these diagrams by specifying the scattering angle $\q$ of $\u{1}-\u{\rm o}$ relative to the inbound direction $\v{\rm o}-\u{\rm o}$ in the initial frame of one of the bodies (the $u$-body $\scS_{u_{\rm o}}$ frame in Fig. \ref{fig:GEN}). This angle specifies the asymptotic direction of the scattering of the $u$-body in the $\scS_{u_{\rm o}}$ frame where it was initially at rest, the reference direction being defined by the CM velocity $\V-\u{\rm o}=\frac{\m{v}}{\m{v}+\m{u}}{\left( \v{\rm o}-\u{\rm o}\right)}$ in that frame. Denoting by $\u{\q}$ the direction defined by $\u{1}-\u{\rm o}$, and $\vb{w}_{\q}$ its orthogonal direction in the plane of the collision, we deduce from the momentum and kinetic energy conservation laws that\citep{RicadaSilva06}
\begin{eqnarray}
\nonumber	&\left\|\u1-\u{{\rm o}}\|\right.= \frac{2  \m{v}}{\m{v}+\m{u}}\left(\v{{\rm o}}-\u{{\rm o}}\right)\cdot\u{\q}\,,\\\\\nonumber
	&\v1-\u{\rm o}=\vb{n}+\u1- \v{{\rm o}},
	\label{eqn:u1uo} 
	\end{eqnarray} 
with
$
	\vb{n}_{\hphantom{1}}=2\, \left\|\v{{\rm o}}-\u{{\rm o}}\right\|\,\sin(\q )\vb{w}_{\q }\;,
$
from which follows
	\begin{eqnarray}
	\u1&=&\u{{\rm o}}+\frac{2  \m{v}}{\m{v}+\m{u}}\left\|\v{{\rm o}}-\u{{\rm o}}\right\|\cos(\q )  \u{\q}\;,\\
	\v1&=&\u{{\rm o}}+\left\|\v{{\rm o}}-\u{{\rm o}}\right\|\,\left( \frac{\m{v}-\m{u}}{\m{v}+\m{u}}\,\cos(\q )\,\u{\q }+\sin(\q )\, \vb{w}_{\q }\right)\;.\;
	\label{eqn:vn}
\end{eqnarray} 
together with the orthogonality equations
\begin{eqnarray}
\nonumber	\vb{n}\cdot \left(\u1-\u{{\rm o}}\right)=0&\,,\quad
	\left(\vb{d}-\left(\u1-\u{{\rm o}}\right)\right)\cdot\left(\u1-\u{{\rm o}}\right)=0\;\,,\\\\
\nonumber	\left(\vb{D}-\vb{n}\right)\cdot\vb{n}=0\;,
\label{eqn:oc3}
\end{eqnarray}
where
\begin{equation}
\vb{d}= 2(\vb{V}_{{\rm cm}}-\u{{\rm o}})= \frac{2  \m{v}}{\m{v}+\m{u}}\left(\v{{\rm o}}-\u{{\rm o}}\right)\;\,,\quad
\vb{D}=2(\v{{\rm o}}-\u{{\rm o}})\;.
\end{equation}
In the interesting limiting case where $\m{v}\gg\m{u}$ and $\v{\rm o}\approx\V\approx\v1$ we can anticipate that the $\v1$ circumference reduces to a point at $\V$, meaning that the $\m{v}$ body motion is practically unaltered by the collision. Then $\vb{d}\approx\vb{D}=2\left( \v{\rm o}-\u{\rm o}\right)$ and, for given magnitudes $v_{\rm o},\, u_{\rm o}$, this is maximized for head-on collisions which provides the greatest magnitude variability for the outbound $\u1$. 
This is the case of planetary  flyby by satellites for attractive orbits, and it is immediately apparent that scattering in the direction of the planet's velocity (if possible) has the potential for more dramatic acceleration or deceleration of the satellite. This is however limited by the fact that planets have large radii (eventually including an atmosphere) and the satellite cannot get closer at periapsis than that radius. Thus some values for $\q$ may be excluded as unpractical.

\section{Determination of Orbital parameters}

The binary elastic collision diagrams such as that shown in Fig. \ref{fig:GEN} provide several simultaneous views of the event, to wit the laboratory, Center of Mass and body frames. The possible velocity outcomes for both masses are parametrized by the angle $\q$ that $\u1-\u{\rm o}$ makes with the reference direction $\v{\rm o}-\u{\rm o}$, which is also the direction of the CM velocity in the $\scS_{u_{\rm o}}$ initial body frame. In  this  frame, where $\m{u}$ was initially at rest, $\q$ is the scattering angle of the mass $\m{u}$ after the collision with an incoming mass $\m{v}$. The range of $\q\in\left[-\frac{\pi}{2},+\frac{\pi}{2} \right]$ encompasses all possible results in a binary elastic collision with given initial velocities. Thus $\q$ works in the $\scS_{u_{\rm o}}$ frame  as the extra parameter needed to determine the outcome of the collision uniquely, a  role that is usually attributed to the `impact parameter' ${b}$ (the distance in reduced-mass-frame between the origin  and the asymptotic line drawn from the incoming body, with direction this body's asymptotic relative velocity). This impact parameter  can only be precisely defined in the reduced-mass frame, which we will assume here to be the instantaneous body-frame $\scS_{v}$ of the mass $\m{v}$ (since we are ultimately interested in exploring all directions for the fly-by about a planet of mass $\m{v}>\m{u}$, the satellite will be henceforth represented by the $u$-body of mass $\m{u}$).\enlargethispage{5\baselineskip}

 In the $\scS_{v}$ body-frame (BF) with axes parallel to those of the laboratory frame (LF), the relative motion of the two masses will appear as that of a single reduced mass $\nobreak{\mu={\m{v}\,\m{u}}/({\m{v}+\m{u}})}$, at the relative position $\r{\mu}^{\dprime}$ of $\m{u}$
\EQ{
\r{\mu}^{\dprime}\equiv\r{u}^{\dprime}=\r{u}-\r{v},
} 
which appears to be moving under interaction forces pointing to a fixed total mass $M=\m{v}+\m{u}$ at the origin, where the other mass ($\m{v}$) is at rest, and this can actually be computed for sufficiently well behaved forces. 

Once the (BF) motion $\r{\mu}^{\dprime}(t)$ is obtained, and assuming that the frame directions are parallel to those of the laboratory frame, we may return to the (LF) description by noting that, in the absence of external forces, the CM motion is uniform and therefore
\
\EQ {
\hspace*{-12pt}
\left\{
\begin{array}{ll}
\r{v}(t)= \R(t)-\dfrac{\m{u} }{M}\r{\mu}^{\dprime}(t)={\R}(0)+\V\,t - \dfrac{\m{u} }{M}\,\r{\mu}^{\dprime}(t)\vspace{1em}\,,\\
\r{u}(t)=\R(t)+\dfrac{\m{v} }{M}\r{\mu}^{\dprime}(t)={\R}(0)+\V\,t  +\dfrac{\m{v} }{M}\,\r{\mu}^{\dprime}(t).
\end{array}
\right.\label{eqn:BFtoLF} 
}
This is an approximation in the real case of planetary flyby because of the gravitational influence of the sun, but for the duration of the encounter, assumed to start and end at the boundaries of the planetary sphere of influence, the effect of the third body is assumed to be negligible. \cite{Barrabes04}\vfill

\subsection{\textsc {View from the {\textit{v}}-body frame $\scS_v$ (reduced mass\, system)}}

If the (BF) reference directions were rotated and scaled with respect to the (LF) directions in such a way that $\r{}=\dsA\cdot\r{}^{\dprime} $,  with  $\dsA\in{\scO}_3$ the rotation matrix, then we should write (\ref{eqn:BFtoLF}) in the form
\
\EQ{
\left\{
\begin{array}{ll}
\r{v}(t)={\R}(0)+\V\,t - \dfrac{\m{u} }{M}\,\dsA\cdot\r{\mu}^{\,\prime\prime}(t)\vspace{1em}\,,\\
\r{u}(t)={\R}(0)+\V\,t  +\dfrac{\m{v} }{M}\,\dsA\cdot\r{\mu}^{\,\prime\prime}(t)\,.
\end{array}
\right.\label{eqn:ABFtoLF}
}

\begin{figure*}[t]
	\hspace{-1em}\includegraphics[scale=.70]{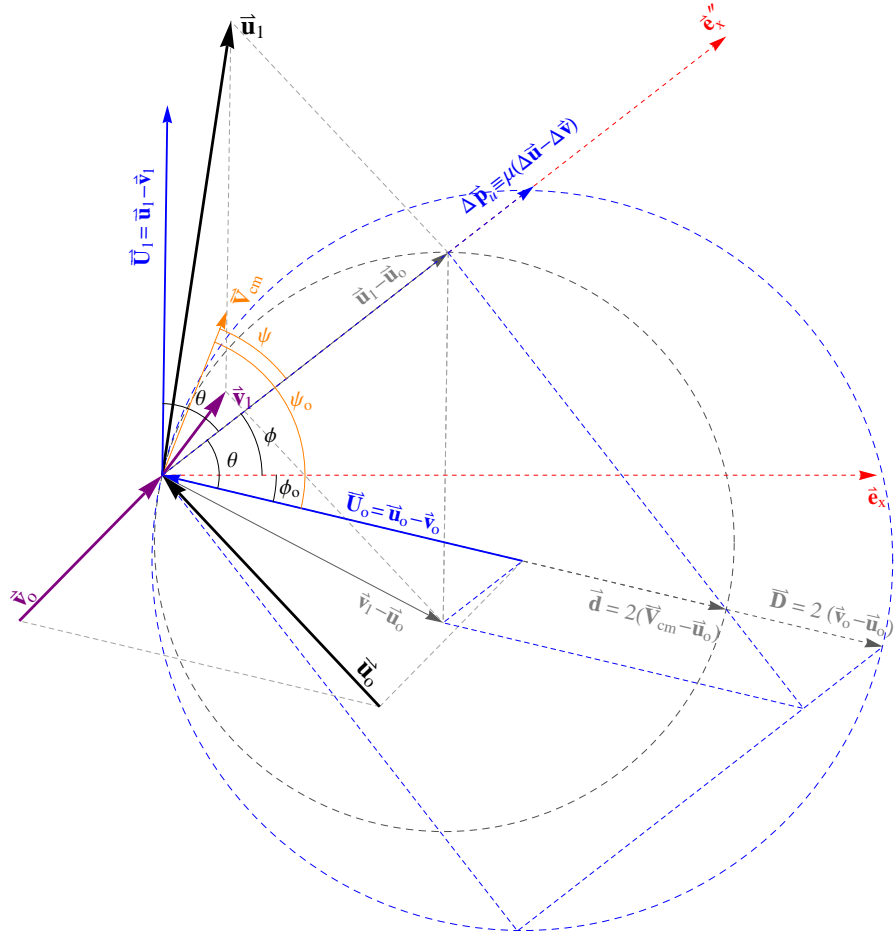}
	\caption{{Asymptotic directions and axis (blue) as viewed in the $v$-body frame. This is a $u$-body slingshot collision where $\m{v}\approx2.6\m{u}$. Notice the interpretation of $\vec{\bold u}_{1}-\vec{\bold u}_{\rm o}$ as proportional to $\Delta\vb{p}_{u}$, the total momentum change of the $u$-body in the reduced-mass $\scS_{v}$ frame, and also as the symmetry axis of the trajectory in the non-inertial $v$-body frame. The angle $\q$ then is identified with the asymptotic angles $\vartheta_{\pm\infty}$ of the trajectory in that frame.}}
	\label{fig:ASYM}\vspace*{0pt}
\end{figure*}

Viewed from the reduced-mass frame ${\scS}_v$ attached to the $v$-body of mass $\m{v}$ with reference directions rotated through $\vb{L}$ by $\dsA^{-1}(\phi)$, the asymptotic incoming velocity $\u{\rm o}^{\dprime}$ of the $u$-body of mass $\m{u}$ obeys (see Fig. \ref{fig:ASYM})
\EQ{
\dsA\cdot\u{\rm o}^{\dprime}=\vb{U}_{\rm o}=\u{\rm o}-\v{\rm o}\,,
}
 which is anti-parallel to the diameter vectors $\vb{d}$ and $\vb{D}$ of the reference circumferences. Likewise, the asymptotic outgoing velocity $\u1^{\dprime}$ for the $u$-body in this frame verifies 
\EQ{
\dsA\cdot\u1^{\dprime}=\vb{U}_{1}=\u1-\v1\,.
} 
On the other hand, in the reduced-mass frame $\scS_{v}$ the mean force acting on the $u$-body during the collision is proportional to that body's total linear momentum variation $\Delta \vb{p}_{\mu}^{\dprime}$ in said frame and
\EQ{
 \dsA\cdot\Delta \vb{p}_{\mu}^{\dprime}=\mu\, \dsA\cdot\left( \u1^{\dprime}-\u{\rm o}^{\dprime}\right)=\mu\,\(\u1-\u{\rm o}+\v{\rm o}-\v1\)\,.
\label{eqn:DP2} 
} 
Since conservation of linear momentum in the laboratory frame  implies
\EQ{\Delta\v{}=\v1-\v{\rm o}=\frac{\m{u}}{\m{v}}\left( \u{\rm o}-\u1\right)=-\frac{\m{u}}{\m{v}}\Delta\u{}\,,} (\ref{eqn:DP2}) is always the sum of collinear terms parallel to $\u1-\u{\rm o}$, and in fact
\EQ{
  \dsA\cdot\Delta \vb{p}_{\mu}^{\dprime}=\mu\,\(\Delta\u{} - \Delta\v{}\)=\m{u}\, (\u1-\u{\rm o})=\Delta\vb{p}_{u}\,.
\label{eqn:DDP} 
} 
In the reduced-mass frame $\scS_v$, for Newtonian or Coulombian type interactions, the asymptotic motion is known to be conic with one focus at the origin\citep{AER05}. In particular for strictly positive total initial energy $\scE^{\dprime}>0$ in the (BF), a gravitational collision must be an hyperbola concave to the focus at the origin, while for repulsive Coulombian forces the hyperbola branch is convex. The direction of $\Delta\vb{p}_{u}$ in the (LF) then also determines the hyperbolic axis $\e{x}^{\dprime}$ for the $u$-body trajectory in ${\scS}_v$, and its angle $\cq_{\mp\infty}$ with the asymptotes $-\u{\rm o}^{\dprime}$ and $\u1^{\dprime}$  is also related to the parameter angle $\q$ (see Figs. \ref{fig:ASYM} and \ref{fig:HCS}). 
Once the angle $\phi$ between the axis $\e{x}^{\dprime}\propto\Delta\vb{p}_{\mu}^{\dprime}$ of the hyperbola and a chosen laboratory frame $\e{x}$ direction is known, the rotation matrix $\dsA(\phi)$ such that $\r{} =\dsA\cdot\r{} ^{\dprime}$ is determined as 
\EQ{
\dsA(\phi)= \csep[1.]{5}\left(
\begin{array}{ccc}
 \cos (\phi ) & -\sin (\phi ) & 0 \\
 \sin (\phi ) & \cos (\phi )  & 0\\
	0     &		0     & 1
\end{array}
\right)\,.
}
assuming that the constant angular momentum is $\vb{L}=L_{z}\e{z}$.

According to the diagrams in Fig. \ref{fig:ASYM}, for a given incoming direction $\u{\rm o}^{\dprime}$ in the $\scS_{v}$ frame, making an angle \EQ{\phi_{{\rm o}}=\arctan{\left[\vbox to 12pt{}\v{\rm o}-\u{\rm o} \right]}} 
with the $\e{x}$ direction in the LF frame, the directions $\phi=\phi_{\rm o}+\q$ of the possible axis $\e{x}^{\dprime}\propto\Delta\vb{p}_{\mu}^{\dprime}$ are in the range
\EQ{
\phi \in \left[\phi_{\rm o}-\frac{\pi }{2},\phi_{\rm o}+\frac{\pi }{2}\right]\,.
} 
The sign of 
${\phi -\phi_{\rm o}=\q}$ is an indication of whether the incoming motion is from above or below the axis 
\EQ{
\e{x}^{\dprime}=\dsA^{-1}(\phi)\cdot\e{x} \,.
}
The angle $\phi=\phi_{\rm o}$ corresponds to the situation where $\e{x}^{\dprime}$ is aligned with $\u{\rm o}^{\dprime}$, i.e. a head-on collision with $\q=0$.

\noindent In the $\scS_v$ frame, the angle $\cq_{-\infty}$ from the $\e{x}^{\dprime}$ axis  to  the asymptotic direction $\nobreak{-\u{\rm o}^{\dprime}=\dsA^{-1}(\phi)\cdot\vb{U}_{\rm o}}$ is ${\cq_{-\infty}=\arctan\left(-\u{\rm o}^{\dprime}\right)=\q}$  and it  determines the boundaries for the actual orbit. For a repulsive interaction, the polar angle $\cq_{\rm r}=\cq$ will change in the interval
${\cq_{\rm r}\in[\cq_{-\infty}\,,-\cq_{-\infty}]}$ while for an attractive interaction $\cq_{\rm a}=\pi-\cq$ and its domain of variation is
${\cq_{\rm a}\in[\pi-\cq_{-\infty}\,, 2\pi-\cq_{-\infty}]}$ as indicated in Fig. \ref{fig:HCS}.

\subsection{\textsc{Initial conditions and determination of orbit parameters}}
\enlargethispage{2\baselineskip}

\begin{figure}
\centering
	\includegraphics[scale=.8]{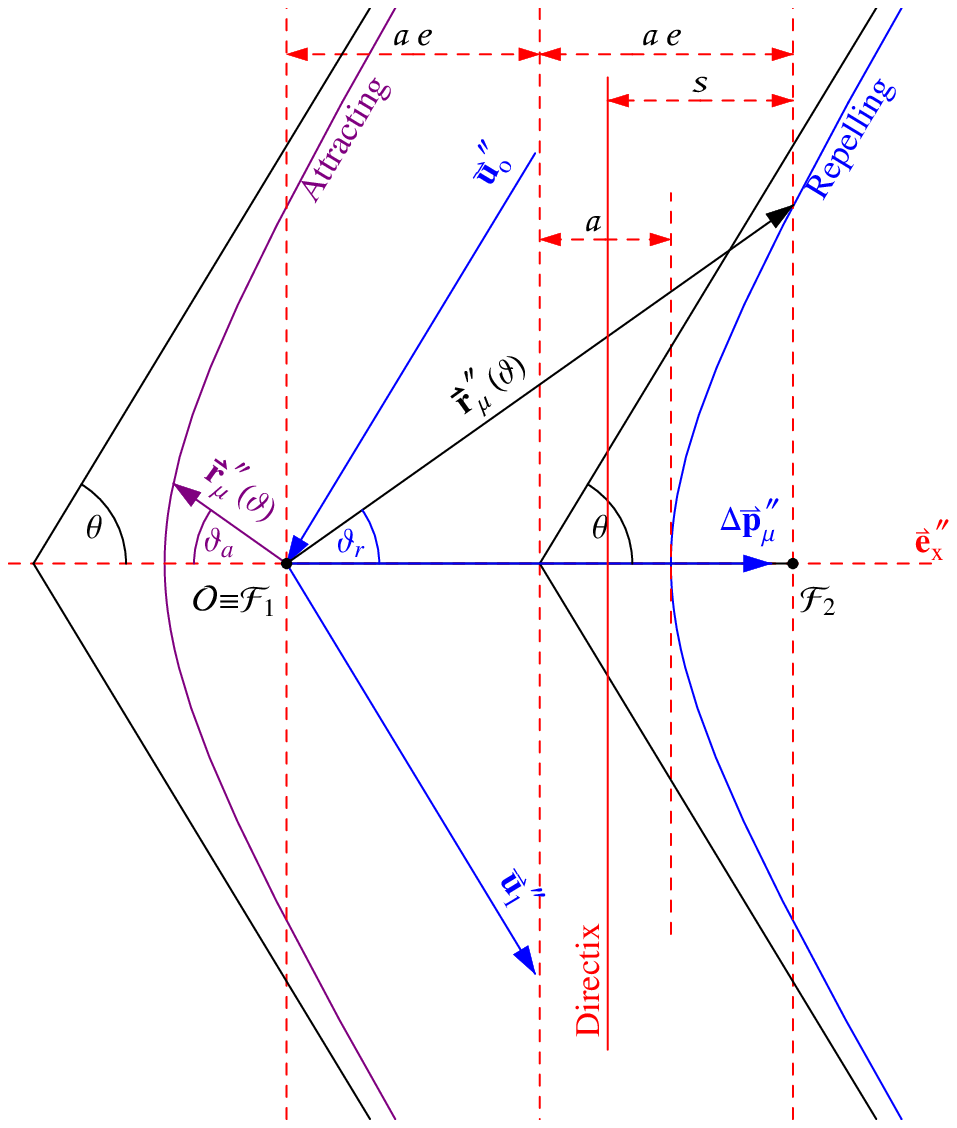}
	\caption{{\hsize=.9\hsize Orbital detail of the attracting and repelling open Keplerian or Coulombian orbits viewed in the  $v$-body frame $\scS_v$. The ${\e{x}}^{\dprime}$ axis is chosen to coincide with the reduced mass total change-of-momentum $\Delta{\vb{p}_{\mu}}^{\dprime}$ in the collision. The asymptotic angle $\q$ is the same as the one identified in figs \ref{fig:GEN} and \ref{fig:ASYM}.}}\vspace{0pt}
	\label{fig:HCS}
\end{figure}

Knowing the asymptotic angle $\cq_{-\infty}=\pi-\q$ of the hyperbolic trajectory $\nobreak{\r{\mu}^{\dprime}(t)\equiv\r{u}^{\dprime}(t)}$ of the reduced mass $\mu$ in the $v$-body frame $\scS_{v}$ determines the eccentricity $\sce$ of this orbit through the relation 
\EQ{\cq_{-\infty}=\cos^{-1}\left(- \frac{1}{\sce}\right)\,.} 
The actual branch of the hyperbola that corresponds to the motion is the one concave towards the focus $\scF_{1}$ at the origin $\scO$ if the interaction is attractive, otherwise it is the one concave to the other focus $\scF_{2}$, situated along the axis $\e{x}^{\dprime}$ at a distance 
\EQ{2\,\scc=2\,\sca\,\sce=\frac{2\, \sce^2\, \scs}{\sce^2-1}\,,} 
from the hyperbolic center. Here $\scs$ is the focal distance (to the directrix) and $\sca$ the semi-major axis (i.e. half the distance between axis intercepts with the hyperbola) see  Fig. \ref{fig:HCS}.

Using polar coordinates $\{\scr^{\dprime},\,\cq\}$ in the $\scS_v$ body frame, an hyperbola  with axis  $\e{x}^{\dprime}$ aligned with  $\dsA^{-1}(\phi)\cdot\dse_{x}$ and a focus at the origin $\scO=\scF_{1}$ will be defined for all eccentricities $\sce>1$ and focal distances $\scs>0$ by the parametric equation
\EQ{\scr_{\mu}^{\dprime}(\cq )=\frac{\sce\, \scs}{1+\, \sce\, \cos (\gamma\pm\cq )}\,,}
where $\gamma$ is a constant dependent of initial conditions. Using henceforth  $\kappa=G\, \m{v}\m{u}$ for a gravitational interaction, or $\kappa=\left|\frac{q_1q_2}{4\,\pi\,\epsilon_{\rm o}}\right|$ for a Coulombian interaction, the time equation can also be expressed as a parametric function of $\cq$ in the form of Kepler's equation
\begin{widetext}
\begin{multline}
\hfil\dst t(\cq)=\dst\left(\frac{\sce \scs}{\sce^2-1}\right)^{3/2} \sqrt{\frac{\mu }{\kappa}} \left(2 \tanh ^{-1}\left[\sqrt{\tst\frac{\sce+1}{\sce-1}} \tan \left(\tst\frac{\cq }{2}\right)\right]
\pm\,\sqrt{\sce^2-1} \,\frac{\sce \, \sin (\cq )}{1\,\pm\,\sce\, \cos (\cq )}\right)\,,\hfil
\end{multline}	
\end{widetext}
where the sign choice distinguishes attractive (+) or repulsive (-) orbits.

From the eccentricity $e$ and the initial energy  in the $\scS_v$ frame, ${\scE_{o }^{\dprime}=\frac{1}{2}\,\mu \,|\u{\rm o}^{\dprime}|^2}$, it is possible to obtain the focal distance $s$  and the angular momentum ${\scL}^{\dprime}$
\begin{eqnarray}
\nonumber &\scs=\dst\frac{ \kappa\,\left(\sce^2-1\right)}{\sce\,\mu\, |\u{\rm o}^{\dprime}|^2}=\frac{\kappa \,\sin(\q )\,\tan(\q )}{\mu \,|\u{\rm o}^{\dprime}|^2}\,,\\\\\nonumber
&{{\scL}^{\dprime}}^{2}=\dst\frac{\,\kappa^{2}\,\left({\sce^2-1}\right)} {|\u{\rm o}^{\dprime}|^{2}} =\frac{\kappa^{2}\,\tan^{2}(\q )} {|\u{\rm o}^{\dprime}|^{2}}\,,
\end{eqnarray}	
and consequently (since ${\scL}^{\dprime}=\mu \,  b\,|\u{\rm o}^{\dprime}|$) the impact parameter $b$  and the displacement $\sca\,\sce$ of each focus from the origin.
\begin{eqnarray}
\nonumber &b =\dst \frac{\kappa  |\tan(\q )|}{\mu\, |\u{\rm o}^{\dprime}|^2}=\frac{\kappa}{\mu\,|U_{\rm o}|^{2}}\sqrt{\frac{1}{\cos(\q)^{2}}-1}\,,\\\\\nonumber
&\sca\,  \sce=\dst \frac{\sce\,\kappa}{\mu\,|\u{\rm o}^{\dprime}|^2}=\frac{\kappa\,\sec(\q )}{\mu\,|\vb{U}_{\rm o}|^2}\,.	
\end{eqnarray}

\section{The Slingshot Manoeuvre}
\enlargethispage{3\baselineskip}

In Fig. \ref{fig:ASYM} we have an example of a collision where the lighter body gains kinetic energy as seen from the laboratory frame. This is a {near-maximum slingshot collision for $u$-body with mass $\m{u}\approx 0.3 \m{v}$ for given initial asymptotic conditions. This is assuming unrestricted periapsis conditions, i.e. point particle collision. In real collisions not all $\q$ angles are accessible in the vicinity of $\frac{1}{2}\psi_{\rm o}$ for the outgoing $u$-body because that would imply a periapsis distance smaller than allowed by the dimensions of the bodies for an elastic collision.}
\

The slingshot manoeuvres are particular cases of the possible outcomes for either an attracting or repelling collision. It is possible to realize them in all collisions with arbitrary mass ratios, even though it only provides significant boosts in cases where a small inertial mass collides with a much larger one moving much faster. The nature of the interaction is irrelevant, as long as it is central and conservative. Evidently we are not considering for the moment variations of the manoeuvre such as aero-gravity assisted slingshots\cite{Armellin06} where at perigee the forces involved are neither conservative nor central. The particular value of $\q$ that corresponds to the theoretical maximum slingshot case is when $\q$ equals half the angle $\psi_{\rm o}$ between $\v{\rm o}-\u{\rm o}$ and $\V$, which means that $\u1$ and $\v1$ would both come out collinear to the CM velocity $\V$. This can be shown as follows. From one of the orthogonality conditions  in equations (\ref{eqn:oc3}) written as
\EQ{\|\u1-\u{\rm o}\|^2=\vb{d}\cdot \left(\u1-\u{\rm o}\right)\,,
}
with 
\EQ{
\vb{d}=2\left(\V-\u{\rm o}\right)=\frac{2\m{v}}{\m{v}+\m{u}}\left(\v{\rm o}-\u{\rm o}\right) \,,
}
one gets after  expanding on both sides, 
\EQ{
\hspace{-1em}{u_1}^2-2\u1\cdot \u{\rm o}+u_{\rm o}^2=2\V\cdot \left(\u1-\u{\rm o}\right)-2\u1\cdot \u{\rm o}+2u_{\rm o}^2 \,.
}
therefore 
\EQ{
\hspace{-1.29em}{u_1}^2=u_{\rm o}^2+2\,V_{\rm cm}\|\u1-\u{\rm o}\|\cos (\psi )\,,
\label{eqn:VUpsi} }
where $\psi $  is the angle from $\u1-\u{\rm o}$ to $\V$. But from the first equation (\ref{eqn:u1uo})
\EQ{
\|\u1-\u{\rm o}\|=\frac{2\m{v}}{\m{v}+\m{u}}\|\v{\rm o}-\u{\rm o}\|\cos (\theta ) \,,
}
where $\theta $  is the angle from $\v{\rm o}-\u{\rm o}$ to $\u1-\u{\rm o}$. Then, if $\psi _{\rm o}$  denotes the angle from $\v{\rm o}-\u{\rm o}$ to $\V$, the relation 
\EQ{\psi =\psi _{\rm o}-\theta  
\label{eqn:psitheta} }
holds for ${ \psi _{\rm o}\in [-\pi, \ \pi ]}$ and $\q\in\left[-\frac{\pi}{2}\,,\,\frac{\pi}{2} \right]$.
Then the magnitude of $\u1$  depends solely on $\theta $ and $\psi_{\rm o}$  through the relations  (\ref{eqn:VUpsi}) to (\ref{eqn:psitheta})
\EQ{
u_{1}^{2}=u_{\rm o}^{2}+\frac{4\m{v}}{\m{v}+\m{u}}V_{\rm cm}\|\v{\rm o}-\u{\rm o}\|\cos (\psi_{\rm o}-\q ) \cos (\theta ) \,.
\label{eqn:uVpq} }

From (\ref{eqn:uVpq}) we conclude that the range of $\q$ that provides for a  $u$-body velocity boost or slingshot (corresponding to $u_{1}^{2}-u_{\rm o}^{2}> 0$) is bounded by boost-break angle 
\EQ{
\q_{\rm bb}=-\tan^{-1}\left[\vbox to 12pt{}\cot(\psi_{\rm o})\right]\,,
}
which is defined through the boost condition 
\EQ{
\cos(\psi_{\rm o}-\q)\cos(\q)> 0\,.
\label{eqn:slc} 
} \enlargethispage{3\baselineskip}
This means that $\m{u}$-boosts will happen for $\q\in\left[\right.-\frac{\pi}{2}\,, \,\q_{\rm bb} \left[ \right.$ if $\psi_{\rm o}<0$ (and for 
$\q \in\left.\right]\q_{\rm bb} \,, \,\frac{\pi}{2}\left.\right]$ if $\psi_{\rm o}>0$).In each case, the remainder of the $\q$-domain will correspond to a breakage of the $u$-body and a boost to the $v$-body velocity. 
From the definition of $\q_{\rm bb}$ we conclude that for $\psi_{\rm o}=\pm\, 0$  all collision results are $\m{u}$-velocity boosts (for instance when $\m{u}<\m{v}$  these are ``head-on'' collision with opposing velocities $|\u{o}|<|\v{o}|$ and arbitrary impact parameter), while for $\psi_{\rm o}=\pm\,\pi$  all scenarios correspond to a $\m{u}$-velocity break (when $\m{u}<\m{v}$ this is a collision where $\m{u}$ ``catches-up'' $\m{v}$ with parallel velocities and arbitrary impact parameter). Not surprisingly, when $\v{\rm o}-\u{\rm o}$ is perpendicular to $\V$ we have equal $\q$ domains for getting a boost or a breakage.

The condition (\ref{eqn:slc}) above also means that $\cos(\psi)>0$, that is, in any slingshot situation we find that the angle $\psi$ from $\u1-\u{\rm o}$ to $\V$ must verify $|\psi|<\frac{\pi}{2}$. Since $\Delta\vb{p}_{u}=\m{u}\,(\u1-\u{\rm o})$, besides representing the total impulse acting on the $u$-body during the maneuver, is also the direction of the force acting on it at the periapsis, the slingshot condition can now be phrased as follows: if at the point of closest encounter the force acting on the $u$-body has a positive component in the direction of the CM velocity, then there will be a boost in the final $u$-body velocity. Otherwise we will obtain a $u$-body velocity break. Notice that this formulation is valid for both gravity assisted and coulombian slingshots, i.e. attractive as well as repulsive interactions.

The angle $\psi_{\rm o}$ can be obtained from initial conditions in terms of the $\v{\rm o}$ to $\u{\rm o}$ angle $\beta_{\rm o}$ (see Fig. \ref{fig:psitheta}), in which case we can express $\q_{\rm bb}$ by
\EQ{\q_{\rm bb}=\tan ^{-1}\left[\frac{(1-\eta )\, \chi_{\rm o}\cos (\beta_{\rm o} )+ \eta \, \chi_{\rm o} ^2-1 }{(1+\eta) \,\chi_{\rm o}  \sin (\beta_{\rm o} )}\right]
\,,\label{eqn:qbb} }
where $\eta=\frac{\m{u}}{\m{v}}$ and $\chi_{\rm o}=\frac{u_{\rm o}}{v_{\rm o}}$.
Notice that when $\eta=1$ and $\chi_{\rm o}=1$ then $\psi_{\rm o}=\frac{\pi}{2}$ for all $\beta_{\rm o}$, and $\q_{\rm bb}=0$, which means that in every collision 
we have equal $\q$ domains for boosting or breaking. For the more common planet-satellite case $\eta\approx 0$ and Eqn. (\ref{eqn:qbb}) reduces to 
\EQ{
 \q_{\rm bb}=\tan ^{-1}\left[\cot (\beta_{\rm o} )-\frac{\csc (\beta_{\rm o} )}{\chi_{\rm o} }\right]\,.
}

\begin{figure*}[t]
\begin{minipage}[b][]{0.45\linewidth} 
\vspace{14pt}\centering
\includegraphics*[scale=.55, bb=0 70 477 377]{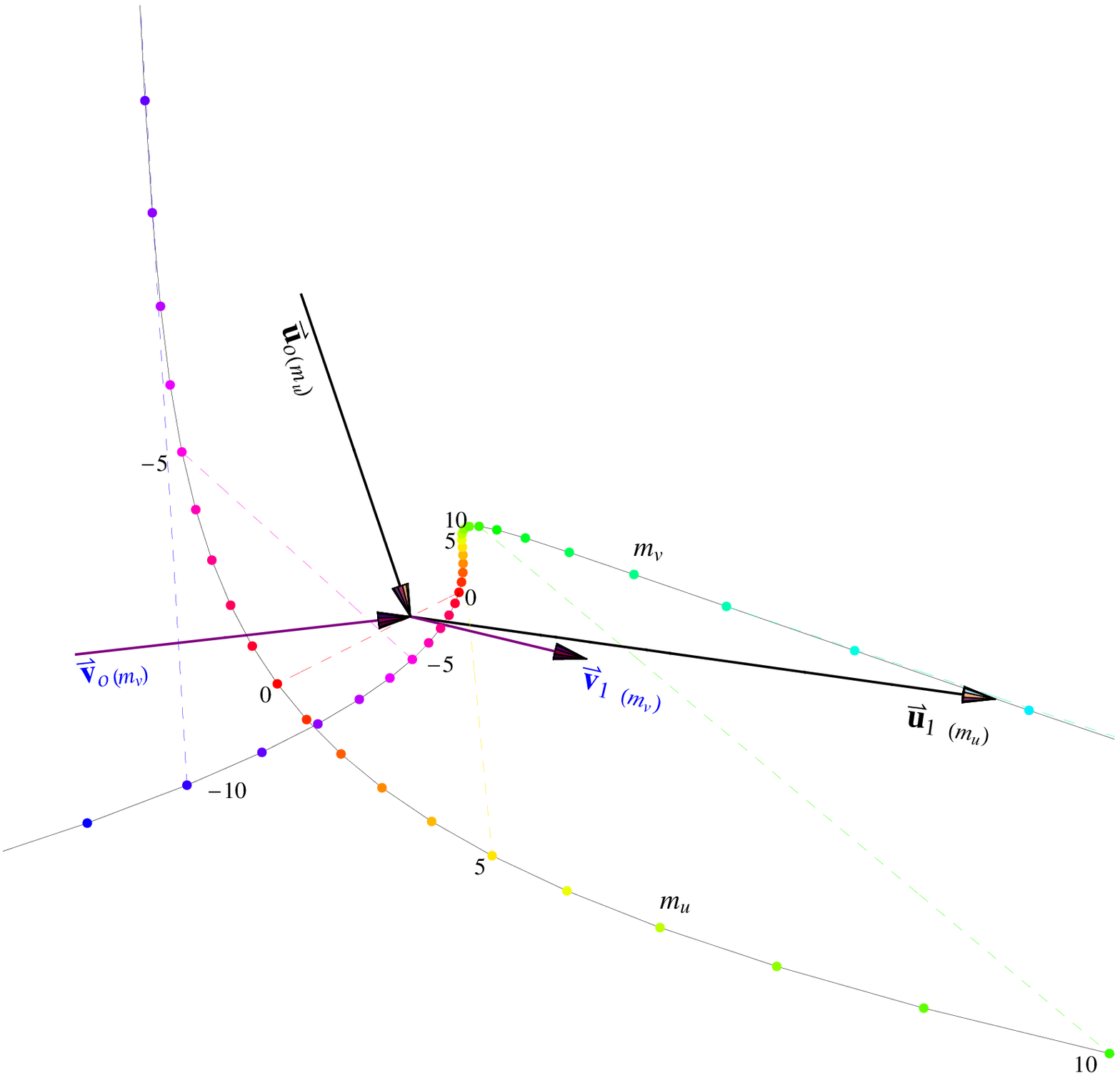}\\
\mbox{(a)}
\hspace{0.5cm} 
\end{minipage}\hspace{1em}
\begin{minipage}[b]{0.45\linewidth}
\centering
\includegraphics*[scale=.55, bb=70 50 427 377]{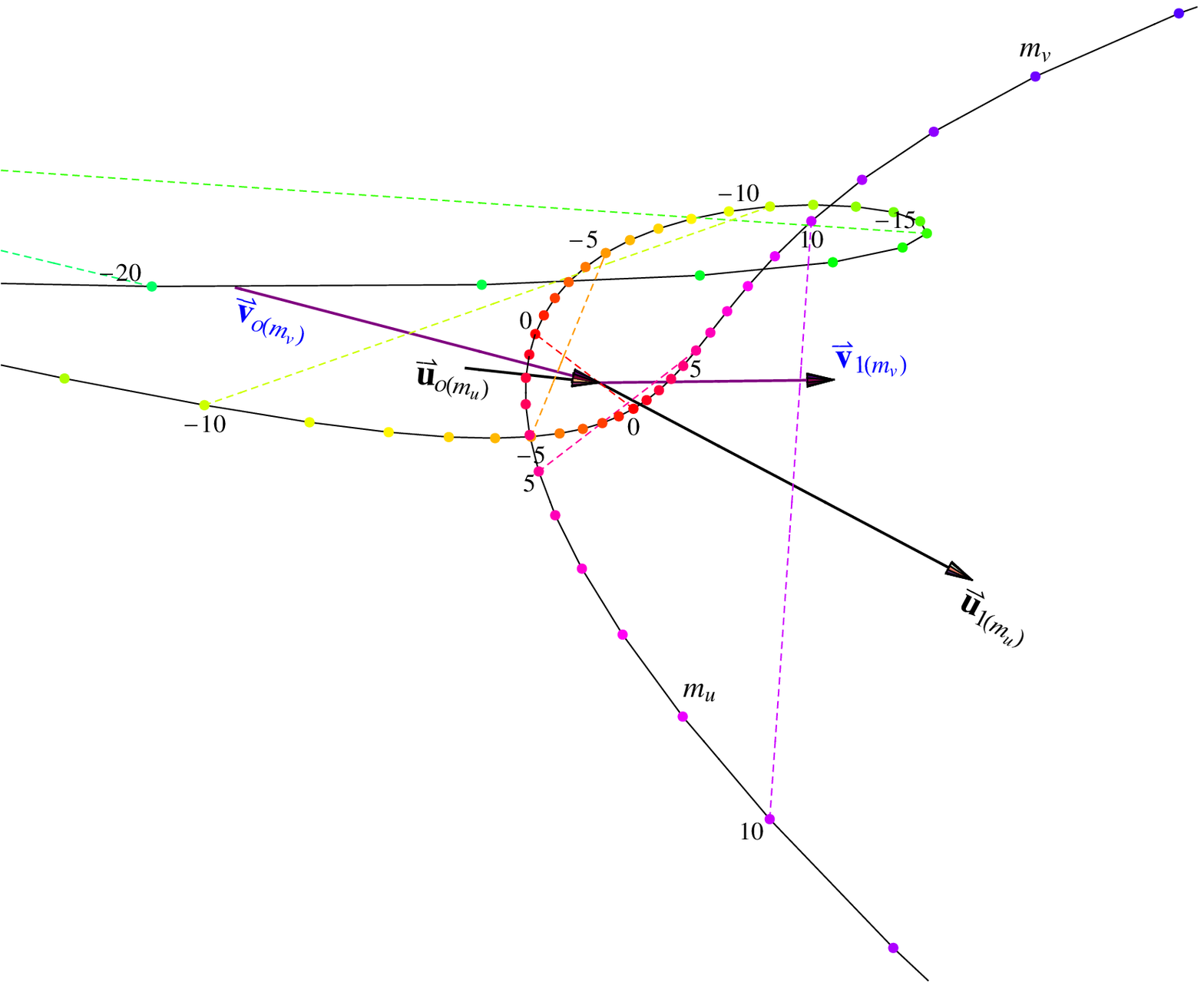}\\
\mbox{(b)}	
\end{minipage}
\begin{minipage}[b]{\linewidth}
\vspace{-12pt}\centering
\includegraphics*[scale=.55, bb=0 100 477 377]{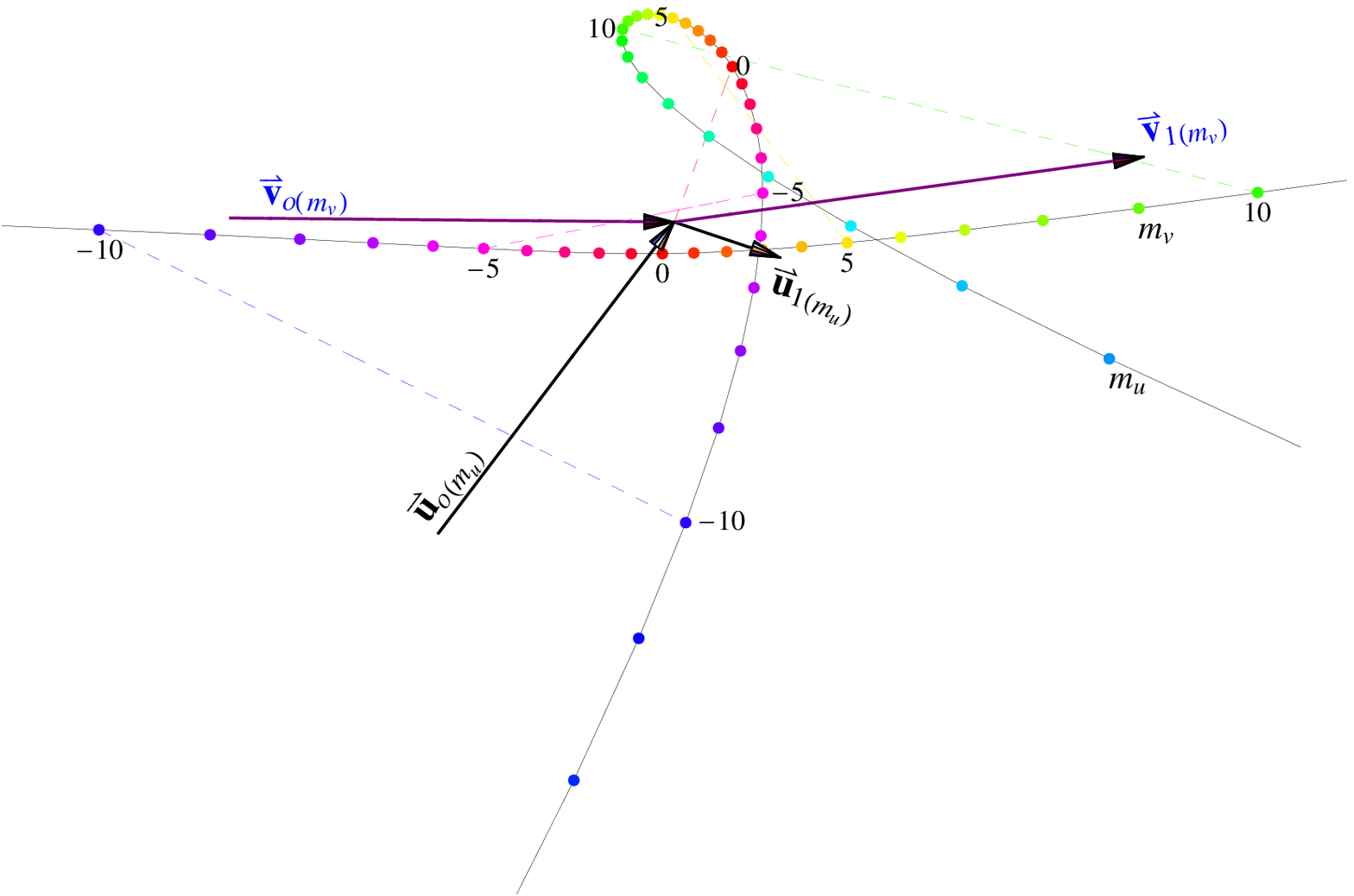}\\
\mbox{(c)}	
\end{minipage}
\caption{{\hsize=.95\hsize  Orbital detail of trajectories for:
(a) an attractive interaction near a maximum slingshot of a $u$-body (black) of mass $\m{u}$ colliding elastically with a mass $\m{v}\approx3\m{u}$. (b) attractive slingshot of a $u$-body (black) of mass $\m{u}$  in a ``catch-up'' collision with a mass $\m{v}\approx2\m{u}$. (c)  breaking maneuver for $u$-body of mass $\m{u}\approx \frac{1}{5}\m{v}$. Color coded dots represent simultaneous (same color) positions in the trajectory,  in equally-spaced $\q$-intervals.  The closest approach corresponds to label $0$. }}
\label{fig:SLCC}\vspace*{22pt}
\end{figure*}

The following figures depict three typical situations in gravity assisted manoeuvres. These were derived from the equations deduced so far for the binary collisions and using diagrams such as those in Figure (\ref{eqn:u1uo}). A live Java applet that models these collisions with a variety of mass ratios and zooming scales can be found in [\cite{AJRS}]. The first two figures correspond to slingshot boosts of the lighter body $m_{u}$, while the last one corresponds to a breaking manoeuvre of $m_{u}$. Figure \ref{fig:SLCC}-(a) represents a slingshot boost for bodies with similar velocities. Figure \ref{fig:SLCC}-(b) is a ``catch-up'' collision where the heavier, faster  body boosts the lighter body velocity as it passes by it. Figure \ref{fig:SLCC}-(c) is a breaking collision for similar velocity bodies. Besides the asymptotic velocities they show the actual trajectories near the periapsis. The color coded points represent same-time position in both orbits. Adjacent point intervals do not correspond to equal time intervals but rather equal angular displacements in the reduced-mass frame orbital representation. The zero index point corresponds to the periapsis. As can be seen from these examples the usual rule-of-thumb for boost or break according to wether the lighter mass passes behind or in front of the heavier one can be very tricky to apply when the mass ratio is not too big. Our proposed formulation above in this section is however rigorous and unambiguous. \enlargethispage{1\baselineskip}

\subsection{\textsc{Slingshot maxima} }
\label{sec:maxima}
\enlargethispage{1\baselineskip}

For ideal point masses the unconstrained extremes are found as usual through 
\EQ{
\d\left(  u_{1}^{2}\right)=\partial_{\psi }\,u_{1}^{2}\,\d\psi +\partial_{\theta }\,u_{1}^{2}\,\d\theta =\left(\partial_{\theta} \,u_{1}^{2}-\partial_{\psi }\,u_{1}^{2}\right)\,\d\theta =0
\,,}
meaning
\EQ{
\sin (\psi ) \cos (\theta )-\cos (\psi ) \sin (\theta )=\sin (\psi -\theta )=0 \,.
}
This holds if $\psi -\theta =n\, \pi $  for integer $n$, but physically only $n=0$  and $n=1$ are of interest. There are thus two extremes. The second variation of $\dst u_{1}^{2}$  defines their type depending on whether $\dst\d^2(u_{1}^{2})\gtrless 0$. But 
\EQ{
\dst\d^2 \left( u_{1}^{2}\right)=\d\sin (\psi -\theta )=-2 \cos (\psi -\theta ) \,\d\theta \,,
} so  $\psi -\theta =0$
  corresponds to a maximum while $\psi -\theta =\pi $  corresponds to a minimum.

Thus the maximum slingshot boost for given incoming initial conditions will happen when $\u1-\u{\rm o}\propto \Delta \vb{p}_2$ makes an equal angle $\psi=\q$ with both $\V$  and $\v{\rm o}-\u{\rm o}$. According to (\ref{eqn:psitheta}) this is
\EQ{
\q_{max}=\frac{1}{2}\,\psi_{\rm o}=\frac{1}{2}\,\arccos\left( \frac{\V\cdot\vb{U}_{\rm o}}{V_{\rm cm}\,U_{\rm o}}\right)\,.
\label{eqn:} 
}
 But this then means that the outgoing asymptotic direction $\u1-\v1$ coincides with $\V$.
Denoting these slingshot extremes by $\vb{u}_1^{\rm sl}$ and $\vb{v}_1^{\rm sl}$ then there is a scalar $\lambda$  such that
 \EQ{{\vb{u}_1}^{\rm sl_{\max }}-{\vb{v}_1^{\rm sl_{\min }}}=\lambda  \vb{V}_{\rm cm} \,.} 
Since in general $\| \vb{u}_1-\vb{v}_1\| =\| \u{\rm o}-\v{\rm o}\| $, the previous equation yields
 \EQ{\lambda =\frac{\| \u{\rm o}-\v{\rm o}\| }{\|\V \|}\,. \label{eqn:slng} } 
From the definition of $\vb{V}_{\rm cm}$ the following holds also in general
 \EQ{\vb{u}_1-\vb{v}_1=\frac{\m{v}+\m{u}}{\m{v}}\left(\vb{u}_1-\vb{V}_{\rm cm}\right)\,, } 
so in conjunction with (\ref{eqn:slng}) 
 \EQ{\frac{\| \u{\rm o}-\v{\rm o}\| }{\|\V \|}\, \vb{V}_{\rm cm}=\frac{\m{v}+\m{u}}{\m{v}}\left(\vb{u}_1^{\rm sl_{\max }}-\vb{V}_{\rm cm}\right)\,, } 
and finally the maximum slingshot velocity is
 \EQ{\vb{u}_1^{\rm sl_{\max }}=\left(1+\frac{\m{v}}{\m{v}+\m{u}}\frac{\| \u{\rm o}-\v{\rm o}\| }{\|\V \|}\right)\vb{V}_{\rm cm} \,.
\label{eqn:MaxU} } \enlargethispage{3\baselineskip}
Corresponding to this maximum, the velocity $\nobreak{\vb{v}_1^{\rm sl_{\min }}=\vb{u}_1^{\rm sl_{\max }}-\lambda  \vb{V}_{\rm cm}}$ will be the minimum of possible $\vb{v}_1$:
\EQ{
\vb{v}_1^{\rm sl_{\min }}=\left(1-\frac{\m{u}}{\m{v}+\m{u}}\frac{\| \u{\rm o}-\v{\rm o}\| }{\|\V \|}\right)\,\vb{V}_{\rm cm}\,. 
} 
On the other hand when $\psi =\theta +\pi$ then $\vb{V}_{\rm cm}$ points in the opposite direction of $\vb{u}_1-\vb{v}_1$, so instead
 \EQ{\vb{v}_1^{\rm sl_{\max }}-\vb{u}_1^{\rm sl_{\min }}=\lambda \vb{V}_{\rm cm} \,,} 
and then
 \EQ{\vb{v}_1^{\rm sl_{\max }}=\left(1+\frac{\m{u}}{\m{v}+\m{u}}\frac{\| \u{\rm o}-\v{\rm o}\| }{\|\V \|}\right)\vb{V}_{\rm cm} \,,} 
 \EQ{\vb{u}_1^{\rm sl_{\min }}=\left(1-\frac{\m{v}}{\m{v}+\m{u}}\frac{\| \u{\rm o}-\v{\rm o}\| }{\|\V \|}\right)\vb{V}_{\rm cm} \,.}

\section{The Constrained Slingshot}

The preceding calculations are however only valid when the periapsis distance of the hyperbolic orbit can be taken as small as necessary, a situation that does not apply in most physical applications. In fact, usually the intervening bodies have dimensions which prevents an elastic collision to occur if the periapsis distance $r_{p}$ is smaller than a limiting value $\scr_{p}<\scr_{\rm min}$. This can be for instance the radius of a planetary mass or star, plus an arbitrary offset to prevent friction from eventual atmospheres or stellar corona. Thus a different maximization must be performed which embodies this constraint when calculating the maximum kinetic energy boost per unit mass $\nobreak{\Delta {\rm k}_{u}=\frac{1}{2} (u_1^{2}-u_{\rm o}^{2})}$ that can be achieved for orbits that do not exceed the limit of maximum approach given by a specific $\scr_{min}$. 

We will first do this calculation in the limiting case of $\V\approx\v{\rm o}$ which is typical for gravity assist flyby of a small satellite about a planetary mass, i.e. when $\m{u}\ll\m{v}$.\newline
Recall that for an hyperbolic orbit the periapsis distance is the $\scr-$value parametrized by $\vartheta=0$, that is 
\EQ{
 \rmin=\frac{\sce\,\scs}{1+\sce}\,.
\label{eqn:periapsis} 
}
The semi-major axis $\sca$ and the focal distance $\scs$ 
\begin{equation}
 \sca=\dfrac{\sce\,\scs}{\sce^{2}-1}\,,\qquad
\scs=\dfrac{\scL^{\dprime}}{\kappa\, \mu\,\sce}\,,
\end{equation} 
can be used to express the kinetic energy $\scE^{\dprime}=\frac{1}{2}\,\mu\,U_{\rm o}^{2}={\kappa}/{2\,\sca}$ and $\nobreak{\rmin=(\sce-1)\,\sca}$ from which follow the identities 
\EQ{\sce=1+\frac{\rmin}{\sca}\,,\qquad\sca=\frac{\kappa}{2\scE^{\dprime}}\,,} where we recall $\kappa=G\,\m{v}\m{u}$.
For an attractive orbit $\cos(\vartheta_{-\infty})=-{1}/{\sce}$ really means $\cos(\q)={1}/{\sce}$ (since $\vartheta_{-\infty}=\pi-\q$) that is
\EQ{
\cos(\q)=\frac{1}{1+\frac{\rmin}{\sca}}=\frac{1}{1+\frac{\mu\,U_{\rm o}^{2}}{\kappa\,/\rmin}}\,.
\label{eqn:cosq} 
} 
\begin{figure*}[t]
	\centering\vspace{-0pt}
	\includegraphics[scale=.83, bb=70 10 439 182]{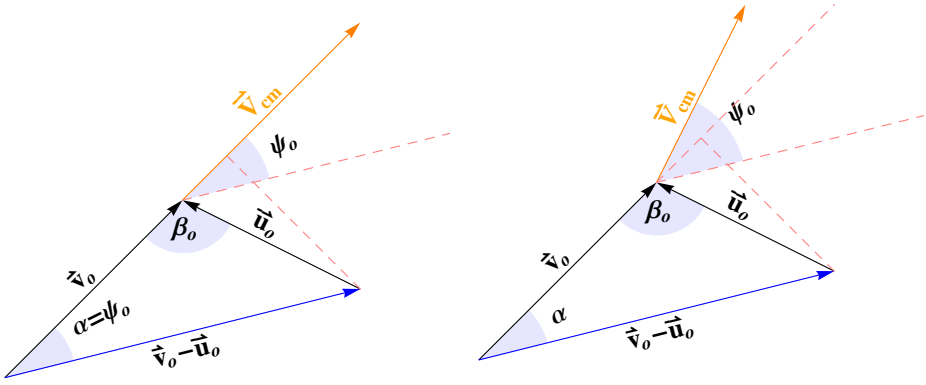}
	\caption{{\hsize=.9\hsize Angular relations in the cases $\m{v}\gg\m{u}$, $\V\approx\v{\rm o}$ (left) and in general (right). Recall that $\psi_{\rm o}=\psi+\q$ in both cases, but $\psi_{\rm o}=\alpha$ on the left, while $\psi_{\rm o}=\alpha+(\psi_{\rm o}-\alpha)$ on the right.}}
	\label{fig:psitheta}
\end{figure*}

Now, since $\V\approx\v{\rm o}$ we obtain the following identities relating initial velocities (see Fig. \ref{fig:psitheta} ):
\begin{equation}
\left\lbrace \begin{array}{ll}
\Uo \sin(\psi_{\rm o})=u_{\rm o}\sin(\beta_{\rm o})\,,\\
\Uo \cos(\psi_{\rm o})=v_{\rm o}-u_{\rm o}\cos(\beta_{\rm o})\,.
\label{eqn:sinpsio} 
\end{array}\right. 
\end{equation}\enlargethispage{3\baselineskip}
where $\beta_{\rm o}$ denotes the incidence angle of $\u{\rm o}$ relative to the direction $\v{\rm o}$ and $\psi_{\rm o}$ is the angle between the directions $\v{\rm o}$ and $\vb{U}_{\rm o}=\u{\rm o}-\v{\rm o}$.
We can use (\ref{eqn:sinpsio}) to eliminate  $U_{\rm o}=\|\v{\rm o}-\u{\rm o}\|$ and $\psi_{\rm o}$ from Eqn. (\ref{eqn:uVpq})  and express the change in kinetic energy per unit mass of the satellite in the laboratory (heliocentric) frame as
\begin{equation}
\hspace{-.8em}\Delta {k}_{u} = \mbox{$\frac{2\, \mu\, V_{\rm cm}\,v_{\rm o}}{\m{u}}$} \cos (\q) ^{2}\left[1 -\mbox{$\frac{u_{\rm o}}{v_{\rm o}}$}\left(\vbox to 11pt{}\cos(\beta_{\rm o})-\sin(\beta_{\rm o})\tan (\q )\right)\right]  \label{eqn:DKu}
\,.\end{equation}
Using Eqn. (\ref{eqn:cosq}) to eliminate $\q$ from Eqn. (\ref{eqn:DKu}) we can express the gain in kinetic energy per unit mass of the satellite in terms of the approach angle $\beta_{\rm o}$ and the velocity ratio $\chi_{\rm o}={u_{\rm o}}/{v_{\rm o}}$.
\begin{widetext}
\begin{equation}
{ \Delta {k}_u}=\frac{2\,\mu\,{v}_{\rm o}^2 }{\m{u}}\,\left[\frac{1-\chi_{\rm o}\,\left(\cos (\beta_{\rm o} )-\sin (\beta_{\rm o} ) \sqrt{\left(1+
\frac{\mu\,\, {v}_{\rm o}^2 }{\kappa \,/\rmin }\left(\chi_{\rm o} ^2-2 \cos (\beta_{\rm o} ) \,\chi_{\rm o} +1\right)\right)^2-1}\right) }{\left(1+\frac{\mu\, {v}_{\rm o}^2 }{\kappa\,/\rmin }\left(\chi_{\rm o} ^2-2 \cos (\beta_{\rm o} )\, \chi_{\rm o} +1\right)\right)^2}\right]
\label{eqn:deltak} \,.
\end{equation}
\end{widetext}

Notice that in (\ref{eqn:deltak}) under the assumed conditions ${\mu}/{\m{u}}\approx 1$ and ${{\kappa}/{\rmin \,\mu}\approx {G\,\m{v}}/{\rmin}}$  is the potential energy per unit mass at the periapsis distance of the planet.

Now we will derive the general relation for the energy gain per unit mass in situations where the momenta of the bodies are of similar order in magnitude, in which case $\V$ is no longer collinear with $\v{\rm o}$. 
In this case the angular relations presented in Fig. \ref{fig:psitheta} indicate that the convenient decomposition for $\psi$ is
\EQ{
 \psi=\psi_{\rm o}-\q  =\alpha+(\psi_{\rm o}-\alpha)-\q \,,
\label{eqn:psioalpha} 
} 
where $(\psi_{\rm o}-\alpha)$ is the angle between $\V$ and $\v{o}$. 
Furthermore, using the definition of $\V$, the projection $\V\cdot\v{\rm o}$ can be expressed as a function of $\beta_{\rm o}$ and the ratio $\chi_{\rm o}={u_{\rm o}}/{v_{\rm o}}$
\EQ{
 V_{\rm cm}{v}_{\rm o}\cos(\psi_{\rm o}-\alpha)=\mu\,{v}_{\rm o}^{2}\left(\frac{1}{\m{u} }+\frac{1}{\m{v} }\chi_{\rm o}\cos(\beta_{\rm o}) \right) 
\label{eqn:Vvo} \,,
}
and also we can get from $\v{o}\times\V$ that
\EQ{
V_{\rm cm}v_{\rm o}\sin(\psi_{\rm o}-\alpha)=\frac{\mu}{\m{v}}v_{\rm o}^{2}\,\chi_{\rm o}\sin(\beta_{\rm o})\,. 
}
Now we should write (see Figure \ref{fig:psitheta})
\begin{equation}
\left\lbrace \begin{array}{ll}
 \Uo\sin(\alpha)=u_{\rm o}\sin(\beta_{\rm o})\,,\\
\Uo\cos(\alpha)=v_{\rm o}-u_{\rm o}\cos(\beta_{\rm o})\,,\\
\Uo^{2}=v_{\rm o}^{2}\left(\chi_{\rm o}^{2}-2\cos(\beta_{\rm o})\,\chi_{\rm o}+1 \right)\,.
\end{array}\right. 
\label{eqn:sinpsio2} 
\end{equation}
Substitution of Eqns. (\ref{eqn:psioalpha}) to (\ref{eqn:sinpsio2}) and (\ref{eqn:cosq}) into formula (\ref{eqn:uVpq}) gives for equal masses $\m{v}=\m{u}=m$
\EQ{
\Delta{k}_{u}=\frac{2\,\mu ^2 v_{\rm o}^2 }{m^2 }\left[\frac{1-\chi_{\rm o} ^2+2 \chi_{\rm o} \sin(\beta_{\rm o}) \scriptstyle\sqrt{\left(1+\frac{\mu\, v_{\rm o}^2 }{\kappa\,/\rmin }\left(\chi_{\rm o} ^2-2 \,\chi_{\rm o}\cos(\beta_{\rm o}) +1\right)\right)^2-1}}{\left( 1+\frac{\mu\, v_{\rm o}^2 }{\kappa\,/\rmin } \left(\chi_{\rm o} ^2-2 \,\chi_{\rm o}\cos(\beta_{\rm o}) +1\right)\right)^{2}} \right]\label{eqn:br2} \,.
}
and in the general case\vfill
\begin{widetext}
\begin{multline}
\Delta {k}_u=\frac{2\mu ^2v_{\rm o}^2 }{\m{u}^2}\left[ \frac{\m{u}}{\m{v}}\frac{1-\chi_{\rm o} ^2+2 \chi_{\rm o} \sin(\beta_{\rm o}) \scriptstyle\sqrt{\left(1+\frac{\mu v_{\rm o}^2 }{\kappa \left/r_p\right.} \left(\chi_{\rm o} ^2-2 \,\chi_{\rm o}\cos(\beta_{\rm o}) +1\right)\right)^2-1}}{\left( 1+\frac{\mu v_{\rm o}^2 }{\kappa \left/r_p\right.} \left(\chi_{\rm o} ^2-2 \,\chi_{\rm o}\cos(\beta_{\rm o}) +1\right)\right)^{2}}\; + \right. \\\left. \frac{\left(\m{v}-\m{u}\right)}{\m{v}}\,\frac{ 1-\chi_{\rm o} \left(\cos(\beta_{\rm o})-\sin(\beta_{\rm o}) \scriptstyle\sqrt{\left(1+\frac{\mu v_{\rm o}^2 }{\kappa \left/r_p\right.} \left(\chi_{\rm o} ^2-2 \,\chi_{\rm o}\cos(\beta_{\rm o}) +1\right)\right)^2-1}\right)}{\left(1+\frac{\mu v_{\rm o}^2 }{\kappa \left/r_p\right.} \left(\chi_{\rm o} ^2-2 \,\chi_{\rm o}\cos(\beta_{\rm o}) +1\right)\right)^{2}}\right]\,.\label{eqn:br3}
\end{multline}	
\end{widetext}

Notice that in these last formulas we have assumed that the periapsis distance is fixed at $\rmin$. What this distance is depends entirely on the particular nature of the collision type. For a slingshot problem of a small satellite about a planet $\rmin$ might be a small multiple of the planet radius, but for a binary star collision it should probably be greater than the Roche limit to avoid the inelastic effects of the gravitational tides between massive extended bodies.

Formulas (\ref{eqn:deltak}), (\ref{eqn:br2}) and (\ref{eqn:br3}) are generalizations to arbitrary masses and orbital conditions of the known energy change expressions for gravitational slingshot such as those found in Broucke's paper\cite{Bro88}.
\enlargethispage{2\baselineskip}

\section{Application to Gravity-Assist Manoeuvres}

\begin{figure}
\begin{minipage}[b][]{0.75\linewidth} 
\hspace{-42pt}\includegraphics*[scale=.75]{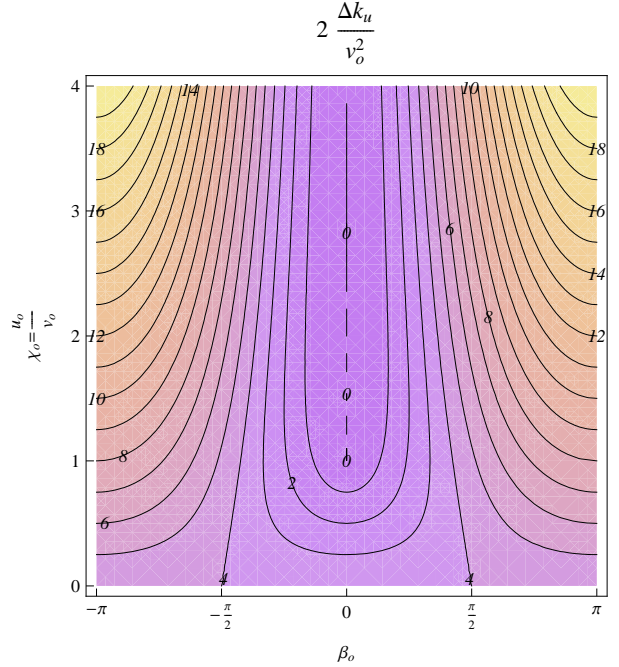}
	\caption{{Boost scenario for a collision with a massive tiny object for which $\frac{R v_{\rm o}^{2}}{G M}\ll 1$.}}
\label{fig:MaxDeltaKU}	
\end{minipage}
\end{figure}

In the particular case of spacecraft manoeuvres assisted by the gravitational field of a planetary object we can assume that $\m{v}\gg\m{u}$ in the formulas above, therefore to a high accuracy $\frac{\mu}{m_{u}}\approx 1$ and $\frac{\mu}{\kappa}\approx \frac{1}{G\,\m{\rm v}}$. When approaching a large planetary mass $M=\m{v}$ from an angle $\beta_{\rm o}$ (measured from $\v{o}$ to $\u{o}$, see Fig. \ref{fig:psitheta}), a satellite of mass $m=\m{u}$ cannot engage the collision with an arbitrary periapsis distance $r_{p}$ and has to maintain a  minimum distance larger than a multiple of the planet's radius $R$. Thus the extreme limiting periapsis distance will be assumed as the planetary radius, $r_{p}\geq R$.  From the expression for the inverse eccentricity $\frac{1}{\sce}=\cos(\q)$ we obtain a limitation on the available orbits by specifying that  possible collision outcomes  must respect the relation
\EQ{
 \cos (\theta )<\frac{1}{1+\frac{ R }{G\,M}\| \u{\rm o}-\v{\rm o}\|^2}\,,
}
or, in terms of the ratio $\chi_{\rm o}=\frac{u_{\rm o}}{v_{\rm o}}$ of satellite to planetary speeds
\EQ{
 \cos (\theta )<\frac{1}{1+\frac{R\,  v_{\rm o}^2}{G\,M}\left(1-2 \cos \left(\beta _{\rm o}\right) \chi _{\rm o}+\chi _{\rm o}^2\right)}\,.
}
This relation limits the maximum velocity outcome at the slingshot through Eqn. \ref{eqn:DKu}, otherwise it could theoretically be 
\EQ{
 u_{1}^{sl_{max}}=v_o \left(1+\sqrt{1-2 \cos \left(\beta _{\rm o}\right) \chi _{\rm o}+\chi _{\rm o}^2}\right)
}
if there were no limitations on the periapsis distance allowed. In terms of relative kinetic energy gain per unit mass we can express this for the theoretical case of an encounter with a massive object with very small radius as 
\EQ{
\frac{2 \Delta{ k}_u}{v_o^2}= \left(1+\sqrt{1-2 \cos \left(\beta _{\rm o}\right) \chi _{\rm o}+\chi _{\rm o}^2}\right){}^2-\chi _{\rm o}^2\,,
}
for which we can expect the general boost pattern as in Figure \ref{fig:MaxDeltaKU}.
For solar system planets however the situation is not as favorable, and we get instead the scenarios shown in Figure \ref{fig:DeltaKU} below.
\begin{figure*}[t]
\begin{minipage}[b][]{0.45\linewidth} 
\vspace{0pt}\centering
\includegraphics*[scale=.75]{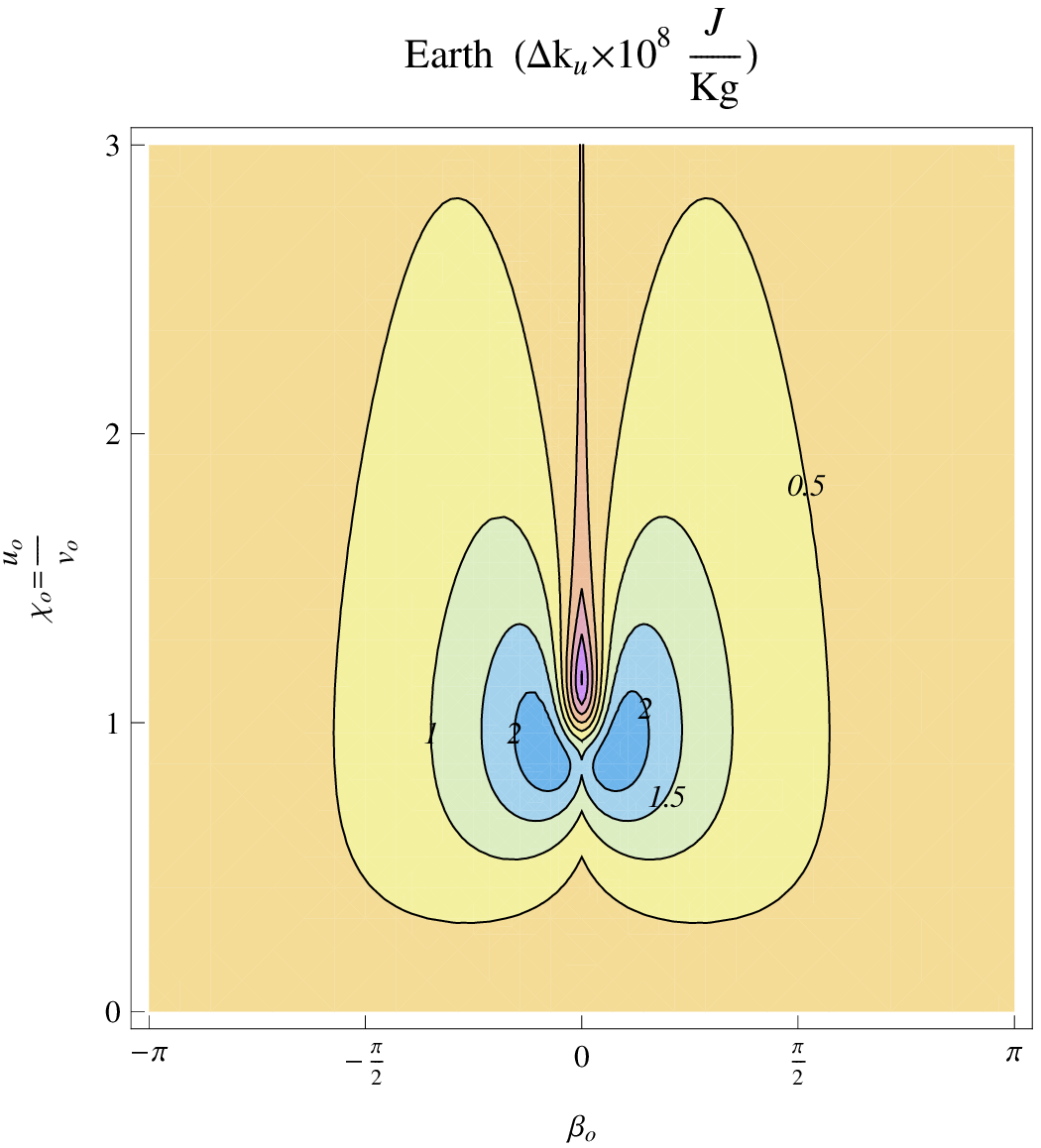}\\
 \mbox{\hspace{7em}  }
\hspace{0.5cm} 
\end{minipage}\hspace{3em}
\begin{minipage}[b]{0.45\linewidth}
\centering
\includegraphics*[scale=.75]{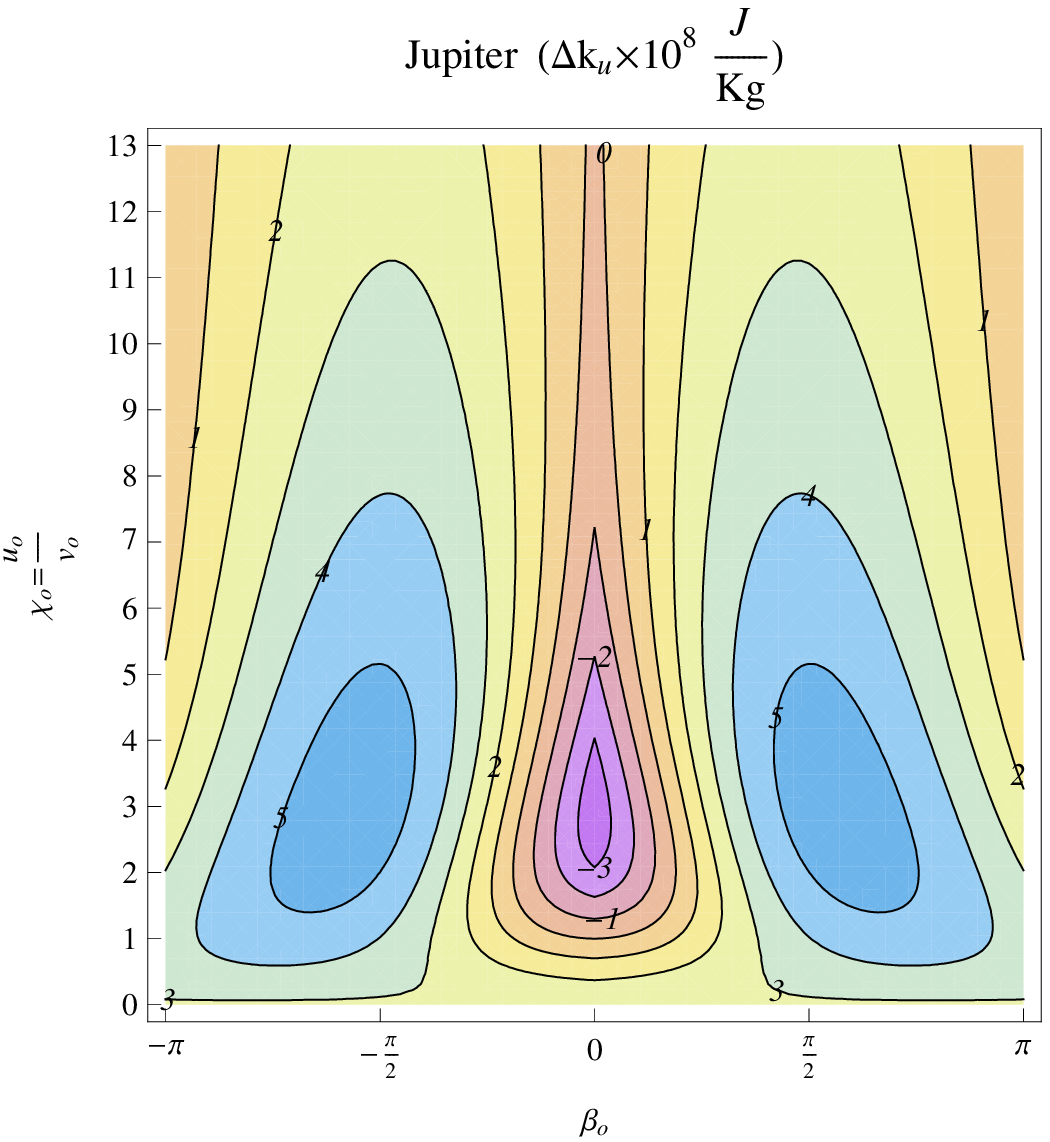}\\
 \mbox{\hspace{5em}   }	
\end{minipage}
\caption{{ Energy gain per unit mass for  Earth and  Jupiter gravity-assist manoeuvres for limiting-case trajectories with periapsis equal to the planetary radius. }\vspace{-4pt}}
\label{fig:DeltaKU}
\end{figure*}
Notice that in both cases the conditions for a breaking manoeuvre are much more limited than those for the boost. For other planets the scenarios are similar to these two cases. These manoeuvres are purely gravitational and do not include the effects of powered boosts at periapsis to modify orbital parameters and thus achieve a different orbit, or aerogravity-assisted manoeuvres to alter the bending angle and yield larger boosts. \cite{Johnson02}

\section{Conclusions}

We have shown that the slingshot effect is no stranger than any other elastic collision, just a particular case of the general set of possible outcomes where the timing of arrival at the point of closest approach is more favorable to a boost in the velocity of the lighter mass. In the case of point particles, equation (\ref{eqn:MaxU}) indicates that the optimal policy is a quasi-head-on for attractive interactions, and a head-on collision for repelling interactions. However, for extended bodies there is a minimum distance $r_{\rm min}$ of approach beyond which there is a severe departure from elasticity (e.g. crash and burn), therefore the approach angle $\q$ for $\u{\rm o}-\v{\rm o}$ must provide for sufficient eccentricity $\sce$ such that $(\sca-1)\,\sce>r_{\rm min}$ for attractive ones, or $(\sca+1)\,\sce>r_{\rm min}$ for repulsive interactions.

In the gravitational case, the Hohmann transfer orbits \citep{WEW97} may be the more energetically efficient but they are not necessarily the optimum policies for  approaching a slingshot configuration. Still in this case there is often the question of whether there could be a solar slingshot manoeuvre. The answer is obviously affirmative. In fact, just like any other elastic binary collision, a slingshot around the Sun is a possibility when viewed from a frame where the Sun itself is moving. Also obvious is nonetheless the fact that, just as in the case of the Jupiter gravity-assist fly-by where there is no gain in the satellite velocity when viewed from the planetary frame, any slingshot manoeuvre involving the sun would appear to return a disappointingly unaltered final speed in the solar reference frame, but this need not be the view from another frame. 
However it is true that for most velocities $\u{\rm o}^{\prime}$ we are able to throw spacecraft with towards the Sun, as seen from that external frame where the Sun itself is moving with velocity $\v{\rm o}\approx\V$ the spacecraft velocity would be 
\EQ{
 \u{\rm o}=\u{\rm o}^{\prime}+\V
}
and thus the boost term in (\ref{eqn:MaxU}) would be 
\EQ{
 \frac{\m{v}}{\m{v}+\m{u}}\frac{\| \u{\rm o}-\v{\rm o}\| }{\|\V \|}\approx \frac{\|\u{\rm o}^{\prime}\|}{\|\V\|}\,.
}
So unless $\|\u{\rm o}^{\prime}\|$ is already much greater than the speed of the solar system itself, this will in general be a small factor and the resulting speed of the spacecraft is about the same as that of the Sun itself. There is also the question of producing a positive energy orbit for a spacecraft that is already within the sphere of influence of the Sun without which the incoming orbit would not be truly hyperbolic.

Note that angular momentum has no role in these calculations, other than being a globally conserved constant that defines the plane where the collision takes place and defines the relation between the impact parameter $\scb$ and the parameter $\q$. So it is surprising to see that many so-called didactic explanations of the effect still mention `stealing angular momentum' to account for the increase in velocity of a spacecraft in a gravity assisted fly-by. It should also be stressed that, although in the specific case of planetary fly-by the calculation of actual orbits involve a complex $N$-body problem, there is no aspect of the slingshot effect above that involve more than two-body interactions. In particular no three-body effects are needed to understand the slingshot of spacecraft in the vicinity of planetary masses \citep{JAVA03}, although the calculations for intermediate trajectories and timings for launch and arrival at the planet vicinity should include this for added accuracy. Tisserand's criteria and three-body graphical methods have been proposed for these high-precision calculations \citep{JKM}, but the fact remains that the slingshot effect involves basically the mechanics of an elastic binary collision.

Having said this, mention should also be made to the limitations on the presented model for the calculations of an actual planetary fly-by. The provisos made in [\citep{RicadaSilva06}] concerning the asymptotic nature of the collision process must be pondered with actual data: the times involved in the approach and extraction of a spacecraft from the fly-by should be compared to the duration of the fly-by before they can be considered as asymptotically infinite. For the duration of the fly-by, the external forces must be negligible as compared to the collisional interaction to introduce only minor perturbations in the resulting trajectories, since it takes too long to disregard the fact that both the planet and the spacecraft are orbiting the Sun. Thirdly, the entry in the planet's `sphere of influence' (which is often considered to be the Hill's sphere) depends on the approach that is made, and that should also be factored in the calculations. Still, as long as these factors can be shown to introduce  small perturbations to the simplified binary collision model, this can be used to successfully explain the physics of the effect. 
\enlargethispage{3\baselineskip}

\bibliographystyle{unsrtnat} 
\bibliography{Orbits}

\end{document}